\documentclass[preprint,showpacs,preprintnumbers,amsmath,amssymb]{revtex4-1}
\usepackage{mathrsfs}
\usepackage{graphicx}
\usepackage{dcolumn}% Align table columns on decimal point
\usepackage{bm}% bold math
\begin{document}

\title{On polarization of vector light beams: origin of Berry phase}

\author{Chun-Fang Li\footnote{Email address: cfli@shu.edu.cn}}

\affiliation{Department of Physics, Shanghai University, 99 Shangda Road, 200444
Shanghai, China}

\date{\today}% It is always \today, today,

\begin{abstract}

When generalized from plane waves to general vector beams, the notion of polarization described by the Stokes parameters turns out to be defined in a momentum-associated system that is fixed by the so-called Stratton vector. As the true intrinsic degree of freedom in the language of quantum mechanics, the polarization of light beams in any fixed momentum-associated system is able to characterize their vectorial feature in the laboratory reference system. The Stratton vector is therefore the degree of freedom to distinguish the vectorial feature of light beams that have the same ``polarization''. Such an observable effect of the Stratton vector helps to understand why plane waves of the same helicity and the same momentum can be different by a Berry phase. This might be the first time to reveal the physical origin of the Berry phase.

\end{abstract}

\pacs{42.25.Ja, 42.90.+m, 03.65.Vf}% PACS, the Physics and Astronomy
%\keywords{}                       % Use showkeys class option if keyword
                                   % display desired
\maketitle

%----------------------------------------------------------------

\section{Introduction}

It is traditionally thought that the polarization of a light beam that is described by the Stokes parameters \cite{BW, Damask} is uniquely determined \cite{JKMW, GJLS}, a conclusion that is drawn basically from a discussion on a plane wave \cite{JR}. As is known, due to the constraint of transversality condition,
$\nabla \cdot \mathfrak{E}=0$,
the electric vector of a plane wave is perpendicular to its momentum. For each momentum there exists a pair of mutually orthogonal base modes.
Consider a monochromatic plane wave that is assumed to propagate along the $z$-axis. If the unit vectors, $\mathbf{e}_x$ and $\mathbf{e}_y$, of the transverse axes are chosen as the electric vectors of the base modes, its electric vector can be expanded as follows,
\begin{equation*}
    \mathfrak{E}^z (\mathbf{x},t)
   =\frac{1}{\sqrt 2}(f_1^z \mathbf{e}_x +f_2^z \mathbf{e}_y)
    e^{i(kz-\omega t)}+\text{c.c.}
\end{equation*}
The complex-valued expansion coefficients $f_1^z$ and $f_2^z$ make up the Jones vector
$\tilde{f}^z=\bigg(\begin{array}{c}
                    f_1^z \\
                    f_2^z
                  \end{array}
             \bigg).
$
The Stokes parameters are determined by the Jones vector as
$\varsigma_i^z=\tilde{f}^{z\dag} \hat{\sigma}_i \tilde{f}^z/\tilde{f}^{z\dag} \tilde{f}^z$
with the following Pauli matrices,
\begin{equation}\label{PM}
    \hat{\sigma}_1=\bigg(\begin{array}{cc}
                           1 &  0 \\
                           0 & -1
                         \end{array}
                   \bigg), \quad
    \hat{\sigma}_2=\bigg(\begin{array}{cc}
                           0 & 1 \\
                           1 & 0
                         \end{array}
                   \bigg), \quad
    \hat{\sigma}_3=\bigg(\begin{array}{cc}
                           0 & -i \\
                           i &  0
                         \end{array}
                   \bigg).
\end{equation}
The problem is that such an expression for the polarization is, strictly speaking, valid only for a plane wave. It is not applicable directly to a general vector beam though it is widely used under the paraxial approximation \cite{Berry, Freu, BBA, Card}. This is because the electric vector of a general beam usually has a non-vanishing component in the propagation direction \cite{LLM, BA, LWY}.
The purpose of this paper is to explore how the expression for the polarization of a plane wave can be generalized to a general beam.
What is found however has significant implications for the intrinsic nature of the polarization.

Now that only the polarization of plane waves is expressible as their Stokes parameters, we consider the plane-wave decomposition of the electric vector of a general beam \cite{CDG, Li09-1, Li16-1},
\begin{equation}\label{FT}
    \mathfrak{E} (\mathbf{x},t)
   =\frac{1}{(2 \pi)^{3/2}} \int \frac{\mathbf{f}(\mathbf{k})}{\sqrt 2}
     e^{i (\mathbf{k} \cdot \mathbf{x}-\omega t)} d^3 k +\text{c.c.},
\end{equation}
where the vector wavefunction $\mathbf{f} (\mathbf{k})$ in momentum space stands for the electric vector of the plane-wave component at the momentum $\mathbf k$. The same as the electric vector of a single plane wave, the electric vector of each of the plane-wave component is perpendicular to its own momentum,
$\mathbf{f} \cdot \mathbf{k}=0$.
This is the expression for the transversality condition in momentum space. But generally speaking, different plane-wave components propagate in different directions and therefore cannot have the same base modes. Let be $\mathbf u$ and $\mathbf v$ the two mutually-perpendicular unit vectors that constitute, with the unit momentum $\mathbf{w}=\frac{\mathbf k}{|\mathbf{k}|}$, a right-handed system $\mathbf{uvw}$, satisfying
\begin{equation}\label{triad}
    \mathbf{u} \times \mathbf{v} =\mathbf{w}, \quad
    \mathbf{v} \times \mathbf{w} =\mathbf{u}, \quad
    \mathbf{w} \times \mathbf{u} =\mathbf{v}.
\end{equation}
Choosing $\mathbf u$ and $\mathbf v$ as the electric vectors of the base modes at the momentum $\mathbf k$, we can expand the electric vector of a general beam as
$\mathbf{f}(\mathbf{k}) =\mathbf{u} f_1 (\mathbf{k}) +\mathbf{v} f_2 (\mathbf{k})$.
The expansion coefficients $f_1$ and $f_2$ make up the corresponding Jones vector
$\tilde{f}(\mathbf{k})=\bigg(\begin{array}{c}
                               f_1 \\
                               f_2
                             \end{array}
                       \bigg).
$
Like the vector wavefunction $\mathbf f$, the two-component Jones vector $\tilde f$ is also a function of the momentum, called the Jones function.
As a result, the Stokes parameters that are determined by the Jones function,
\begin{equation}\label{SP}
    \varsigma_i (\mathbf{k})
   =\frac{\tilde{f}^\dag \hat{\sigma}_i \tilde{f}}{\tilde{f}^\dag \tilde{f}},
\end{equation}
are in general dependent on the momentum.

The key point here is that Eqs. (\ref{triad}) cannot completely fix the transverse axes, $\mathbf u$ and $\mathbf v$, of the local momentum-associated system (MAS) $\mathbf{uvw}$ up to a rotation about the momentum \cite{MW}. This means that for a given light beam $\mathbf f$, the Jones function and hence the Stokes parameters are not unique. In other words, \emph{the polarization of a given light beam is not unique}. To determine the polarization, we have to figure out a way to fix the transverse axes of the MAS $\mathbf{uvw}$, or equivalently, to fix the MAS $\mathbf{uvw}$ itself for all the momenta simultaneously.
Fortunately, it was shown a long time ago by Stratton \cite{Stra} and later on by others \cite{GW, PA, DP} that this can be done by introducing a constant unit vector $\mathbf I$, called the Stratton vector (SV), in the following way,
\begin{equation}\label{axes}
    \mathbf{u} =\mathbf{v}               \times \frac{\mathbf k}{k},          \quad
    \mathbf{v} =\frac{\mathbf{I} \times \mathbf{k}}{|\mathbf{I} \times \mathbf{k}|}.
\end{equation}
The polarization that follows is certainly dependent on the choice of the SV. In this paper we will be concerned with the physical interpretation of the SV-dependent polarization, its role in characterizing the feature of light beams, and the physical significance of the SV.
The contents are arranged as follows.

It is expounded in Section \ref{locality} that the SV-dependent polarization of a light beam is just its property in the local MAS $\mathbf{uvw}$. As the polarization vector, the so-called Poincar\'{e} vector \cite{Merz} in the MAS is introduced with the Stokes parameters as the Cartesian components along the axes of the MAS. It is found that the Poincar\'{e} vectors of a given light beam in different MAS's are generally different. Their relation is also given.
However, the polarization of light beams in any fixed MAS is able to characterize their vectorial feature in the laboratory reference system. The reason is due to the fact that only in one particular MAS can the polarization be, in the language of quantum mechanics, the true intrinsic degree of freedom of the photon. This is discussed in Section \ref{V-feature} in which the concrete meaning of the vectorial feature is defined.
The physical significance of the SV that fixes the MAS is investigated in Section \ref{effect}. Section \ref{remark} concludes the paper with remarks.

\section{Polarization is a property in the MAS}\label{locality}

\subsection{Dependence of the Stokes parameters on the SV}

According to Eq. (\ref{SP}), the Stokes parameters are completely determined by the Jones function. To see what the SV-dependence of the Stokes parameters means, let us first examine how the Jones function is related to the vector wavefunction and how it depends on the SV.
For this purpose, we make use of the Jones function to rewrite the vector wavefunction as follows \cite{Li09-1, Li07, Li2008, Li09-2},
\begin{equation}\label{QUT1}
    \mathbf{f}(\mathbf{k})
   =\varpi \tilde{f} (\mathbf{k}),
\end{equation}
where
$
\varpi=\big(\begin{array}{cc}
              \mathbf{u} & \mathbf{v} \\
            \end{array}
       \big)
$
is a 3-by-2 matrix consisting of the electric vectors of the base modes and vectors of three Cartesian components such as $\mathbf u$ and $\mathbf v$ are expressed as column matrices.
The matrix $\varpi$ depends on the SV via Eqs. (\ref{axes}). However, it is not difficult to prove
\begin{equation}\label{unitarity1}
    \varpi^{\dag} \varpi =I_2,
\end{equation}
irrespectively of the SV, where $I_2$ is the 2-by-2 unit matrix.
Multiplying both sides of Eq. (\ref{QUT1}) by $\varpi^{\dag}$ from the left and considering Eq. (\ref{unitarity1}), we have
\begin{equation}\label{QUT2}
    \tilde{f}(\mathbf{k})=\varpi^{\dag} \mathbf{f}(\mathbf{k}).
\end{equation}

Now we change the SV from $\mathbf I$ to a different one, $\mathbf{I}'$ say. In this case, the Jones function of the same light beam $\mathbf f$ is given by
\begin{equation}\label{QUT2'}
    \tilde{f}'(\mathbf{k})=\varpi'^{\dag} \mathbf{f}(\mathbf{k}),
\end{equation}
where
$\varpi'=(\begin{array}{cc}
            \mathbf{u}' & \mathbf{v}'
          \end{array}
         )
$
and
\begin{equation*}
    \mathbf{u}' =\mathbf{v}' \times \frac{\mathbf k}{k}, \hspace{5pt}
    \mathbf{v}' =\frac{\mathbf{I}' \times \mathbf{k}}{|\mathbf{I}' \times \mathbf{k}|}.
\end{equation*}
Correspondingly, the Stokes parameters are given by
\begin{equation}\label{SP'}
    \varsigma'_i (\mathbf{k})
   =\frac{\tilde{f}'^\dag \hat{\sigma}_i \tilde{f}'}{\tilde{f}'^\dag \tilde{f}'}
\end{equation}
in accordance with the definition (\ref{SP}).
As mentioned above, the transverse axes
$\mathbf{u}'$ and $\mathbf{v}'$
of the primed MAS $\mathbf{I}'$ are related to the transverse axes $\mathbf{u}$ and $\mathbf{v}$ of the unprimed one $\mathbf I$ by a rotation about $\mathbf k$.
Denoted by $\Phi(\mathbf{k}; \mathbf{I}', \mathbf{I})$, the rotation angle is determined by
\begin{equation}\label{varpi'}
    \varpi'=\varpi \exp \left(-i \hat{\sigma}_3 \Phi \right)
\end{equation}
or by
\begin{equation}\label{Varpi'}
    \varpi'=\exp [-i (\hat{\mathbf \Sigma} \cdot \mathbf{w}) \Phi] \varpi,
\end{equation}
where
$(\hat{\Sigma}_k)_{ij} =-i \epsilon_{ijk}$
with $\epsilon_{ijk}$ the Levi-Civit\'{a} pseudotensor.
It is noted that the rotation angle $\Phi$ always depends on the momentum, unless $\mathbf{I}' =-\mathbf{I}$. In that case, we have $\Phi=\pi$.
Substituting Eq. (\ref{varpi'}) into Eq. (\ref{QUT2'}) and noticing Eq. (\ref{QUT2}), we have
\begin{equation}\label{GT-f}
    \tilde{f}'=\exp \left(i \hat{\sigma}_3 \Phi \right) \tilde{f}.
\end{equation}
This is the transformation law of the Jones function under the change of the SV.

The SV-dependence of the Jones function indicates that the Stokes parameters cannot be completely determined by the vector wavefunction.
Substituting Eq. (\ref{GT-f}) into Eq. (\ref{SP'}) and using Eq. (\ref{SP}), we get
\begin{subequations}\label{SP'-SP}
\begin{align}
  \varsigma'_1 & = \varsigma_1 \cos 2\Phi+\varsigma_2 \sin 2\Phi, \label{SP1} \\
  \varsigma'_2 & =-\varsigma_1 \sin 2\Phi+\varsigma_2 \cos 2\Phi, \label{SP2} \\
  \varsigma'_3 & = \varsigma_3.                                   \label{SP3}
\end{align}
\end{subequations}
These relations constitute the transformation law of the Stokes parameters under the change of the SV. They show that the first two Stokes parameters depend on the choice of the SV. Only the third one does not.
On the basis of the transformation laws (\ref{GT-f}) and (\ref{SP'-SP}), we try to interpret the physical meaning of the Stokes parameters.

\subsection{Introduction of the Poincar\'{e} vector in the MAS}

From the fact that the SV plays the role of fixing the MAS it follows that the Jones function should not be defined in the laboratory reference system in which the vector wavefunction is defined. It should be defined in the MAS. In particular, the two different Jones functions (\ref{QUT2}) and (\ref{QUT2'}) are not defined in the same MAS. The former is defined in the unprimed MAS $\mathbf I$. The latter is defined in the primed MAS $\mathbf{I}'$.
By this it is meant that the Stokes parameters determined by the Jones function in any fixed MAS constitute the Cartesian components of a unit vector in that MAS. According to Merzbacher \cite{Merz}, we will call such a unit vector the Poincar\'{e} vector.
Since half the third Pauli matrix is the generator of the transformation (\ref{GT-f}), which comes from the local rotation of the MAS about the momentum, the transformation (\ref{SP'-SP}) suggests that the third Stokes parameter constitutes the longitudinal component of the Poincar\'{e} vector and the first two Stokes parameters constitute the mutually perpendicular transverse components of the Poincar\'{e} vector. Specifically, the Stokes parameters (\ref{SP}) determined by the Jones function (\ref{QUT2}) form the following Poincar\'{e} vector in the MAS $\mathbf I$,
\begin{equation}\label{PV}
    \boldsymbol{\varsigma}
   =\varsigma_1 \mathbf{u} +\varsigma_2 \mathbf{v} +\varsigma_3 \mathbf{w}.
\end{equation}
Likewise, the Stokes parameters (\ref{SP'}) determined by the Jones function (\ref{QUT2'}) form the following Poincar\'{e} vector in the MAS $\mathbf{I}'$,
\begin{equation}\label{PV'}
    \boldsymbol{\varsigma}'
   =\varsigma'_1 \mathbf{u}' +\varsigma'_2 \mathbf{v}' +\varsigma'_3 \mathbf{w}.
\end{equation}
Substituting Eqs. (\ref{SP'-SP}) into Eq. (\ref{PV'}) and taking Eqs. (\ref{varpi'}) and (\ref{PV}) into account, we find
\begin{equation}\label{GT-PV}
    \boldsymbol{\varsigma}'
   =\exp{[i(\hat{\mathbf \Sigma} \cdot \mathbf{w})\Phi]} \boldsymbol{\varsigma}.
\end{equation}
Let us discuss the meaning of this result with the transverse components of the Poincar\'{e} vectors at one particular momentum.

Eqs. (\ref{SP1}) and (\ref{SP2}) state that if the Stokes parameters $\varsigma'_1$ and $\varsigma'_2$ were specified in the MAS $\mathbf I$ in which the Stokes parameters $\varsigma_1$ and $\varsigma_2$ are specified, they would form a transverse component,
$\boldsymbol{\varsigma}''_\perp=\varsigma'_1 \mathbf{u}+\varsigma'_2\mathbf{v}$,
that is equal to the result of the rotation of the transverse component
$\boldsymbol{\varsigma}_\perp=\varsigma_1 \mathbf{u} +\varsigma_2 \mathbf{v}$
by an angle $-2 \Phi$ as is indicated in Fig. 1(a).
But as just mentioned, they should be specified in the MAS $\mathbf{I}'$. So along with the MAS $\mathbf I$ being rotated to the MAS $\mathbf{I}'$ by an angle $\Phi$, $\boldsymbol{\varsigma}''_\perp$ is rotated to
$\boldsymbol{\varsigma}'_\perp=\varsigma'_1 \mathbf{u}'+\varsigma'_2 \mathbf{v}'$ as is indicated in Fig. 1(b).
$\boldsymbol{\varsigma}'_\perp$ is thus equal to the result of the rotation of $\boldsymbol{\varsigma}_\perp$ by an angle $-\Phi$. This is what Eq. (\ref{GT-PV}) means. It reveals that the Poincar\'{e} vectors (\ref{PV}) and (\ref{PV'}) in different MAS's are generally different.
From this result it is concluded that what the Stokes parameters describe is just a property of the light beam in the MAS. In other words, the notion of polarization is a SV-dependent local property of the light beam. Hereafter we will refer only to the Poincar\'{e} vector (\ref{PV}) in the MAS as the polarization vector.
\begin{figure}[tb]
\centerline{\includegraphics[width=11cm]{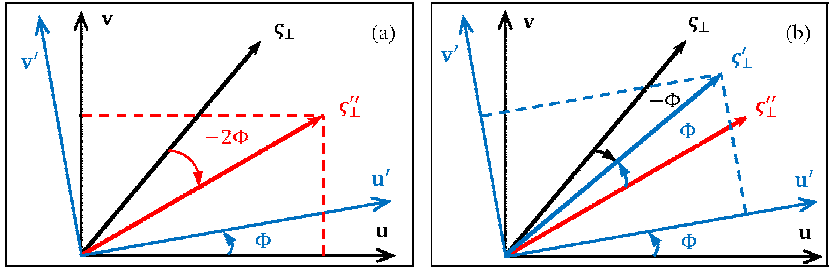}}
\caption{(a) $\boldsymbol{\varsigma}''_\perp$ is equal to the result of the rotation of $\boldsymbol{\varsigma}_\perp$ by an angle $-2 \Phi$. (b) $\boldsymbol{\varsigma}''_\perp$ is rotated to $\boldsymbol{\varsigma}'_\perp$ along with the MAS $\mathbf I$ being rotated to the MAS $\mathbf{I}'$ by an angle $\Phi$.}
\end{figure}

It is now clear that the polarization as a SV-dependent quantity cannot be completely determined by the vector wavefunction. But we will see that the polarization of light beams in any particular MAS is able to characterize their vectorial feature in the laboratory reference system.

\section{The role of the polarization and its nature}\label{V-feature}

\subsection{Characterization of the vectorial feature in the laboratory reference system}

Eq. (\ref{SP}) indicates that the Stokes parameters have nothing to do with the norm of the Jones function. To exploit this property, we write the Jones function as
\begin{equation}\label{tildef}
    \tilde{f} (\mathbf{k})
   =\tilde{a} (\mathbf{k}) f (\mathbf{k})
\end{equation}
to introduce the unit Jones function $\tilde{a}$ that satisfies
$\tilde{a}^\dag \tilde{a}=1$.
On one hand, Eq. (\ref{tildef}) allows us to express the polarization vector (\ref{PV}) in terms of the unit Jones function as follows,
\begin{equation}\label{PV-a}
    \boldsymbol{\varsigma}
   =\tilde{a}^\dag \hat{\boldsymbol \sigma} \tilde{a},
\end{equation}
where
$\hat{\boldsymbol \sigma}
=\hat{\sigma}_1 \mathbf{u}+\hat{\sigma}_2 \mathbf{v}+\hat{\sigma}_3 \mathbf{w}$.
It means that the polarization in the MAS is fully represented by the Pauli matrices (\ref{PM}) in the sense that its Cartesian components in the MAS are given by
$\varsigma_i=\tilde{a}^\dag \hat{\sigma}_i \tilde{a}$.
On the other hand, substituting Eq. (\ref{tildef}) into Eq. (\ref{QUT1}), we have
\begin{equation}\label{vecf}
    \mathbf{f}(\mathbf{k})=\mathbf{a}(\mathbf{k}) f(\mathbf{k}),
\end{equation}
where
\begin{equation}\label{QUT1-a}
    \mathbf{a}(\mathbf{k})=\varpi \tilde{a}(\mathbf{k})
\end{equation}
is the unit vector wavefunction satisfying
\begin{equation}\label{unitarity}
    \mathbf{a}^\dag \mathbf{a}=\tilde{a}^\dag \tilde{a}
\end{equation}
by virtue of Eq. (\ref{unitarity1}).
The same as the vector wavefunction itself, the unit vector wavefunction is also defined in the laboratory reference system and is constrained by the transversality condition,
\begin{equation}\label{TC-a}
    \mathbf{a} \cdot \mathbf{k}=0.
\end{equation}
Multiplying both sides of Eq. (\ref{QUT1-a}) by $\varpi^{\dag}$ from the left and taking Eq. (\ref{unitarity1}) into account, we have
\begin{equation}\label{QUT2a}
    \tilde{a}=\varpi^\dag \mathbf{a}.
\end{equation}
This shows that the unit Jones function does not depend on the norm of the vector wavefunction. It depends only on the unit vector wavefunction.

The matrix $\varpi$ in Eqs. (\ref{QUT1-a}) and (\ref{QUT2a}) performs a quasi unitary transformation in the following sense. Eq. (\ref{QUT1-a}) says that the matrix $\varpi$ acts on a unit Jones function to yield a unit vector wavefunction. Meanwhile, Eq. (\ref{QUT2a}) says that the matrix $\varpi^{\dag}$ acts on a unit vector wavefunction to yield a unit Jones function. Substituting Eq. (\ref{QUT2a}) into the right-handed side of Eq. (\ref{unitarity}) and considering the arbitrariness of $\mathbf a$, we get
\begin{equation}\label{unitarity2}
    \varpi \varpi^{\dag} =I_3,
\end{equation}
irrespectively of the SV. Eqs. (\ref{unitarity1}) and (\ref{unitarity2}) express the quasi unitarity \cite{Golub} of the transformation matrix $\varpi$. $\varpi^{\dag}$ is the Moore-Penrose pseudo inverse of $\varpi$, and vice versa.
Now that the quasi unitarity of $\varpi$ has nothing to do with the SV, to each unit vector wavefunction there corresponds a unique unit Jones function via Eq. (\ref{QUT2a}) and hence a unique polarization vector via Eq. (\ref{PV-a}) once the MAS and the associated matrix $\varpi$ are fixed by any particular SV.
By this it is meant that the polarization in any fixed MAS is able to characterize the feature that is described by the unit vector wavefunction in the laboratory reference system.
Considering that what makes the electric field (\ref{FT}) a vector field, which is experimentally observable, is the very unit vector wavefunction in accordance with Eq. (\ref{vecf}), such a feature will be called the vectorial feature.
The ability for the polarization to characterize the vectorial feature in the laboratory reference system lies with its intrinsic nature in the MAS from the standpoint of quantum mechanics.

\subsection{Intrinsic nature}

The transversality condition (\ref{TC-a}) on the unit vector wavefunction states that the vectorial feature in the laboratory reference system cannot be independent of the momentum.
However, because no conditions such as Eq. (\ref{TC-a}) exist for the Jones function, the Stokes parameters in any MAS can be independent of the momentum. In fact, it is seen from Eq. (\ref{PV-a}) that if the unit Jones function in some MAS is constant, so are the Stokes parameters in that MAS.
What is peculiar here is that the Stokes parameters of a light beam, which are constant in one MAS, are generally not constant in others.

To see this in more detail, we consider a light beam the Stokes parameters $n_i$ of which in the MAS $\mathbf I$ are constant, forming the polarization vector
$\mathbf{n}=n_1 \mathbf{u} +n_2 \mathbf{v} +n_3 \mathbf{w}$.
If the transverse Stokes parameters $n_1$ and $n_2$ do not vanish simultaneously, it follows from Eqs. (\ref{SP1}) and (\ref{SP2}) that the transverse Stokes parameters in any other MAS $\mathbf{I}'$ are no longer constant, unless $\mathbf{I}'=-\mathbf{I}$.
On the other hand, if $n_1$ and $n_2$ are both equal to zero, the unit Jones function of the light beam in the MAS $\mathbf I$ is actually the eigen function of the third Pauli matrix $\hat{\sigma}_3$,
$
\tilde{a}_{\sigma_3}=\frac{1}{\sqrt 2} \bigg(\begin{array}{c}
                                               1 \\
                                               i \sigma_3
                                             \end{array}
                                       \bigg),
$
which satisfies
\begin{equation}\label{EE}
    \hat{\sigma}_3 \tilde{a}_{\sigma_3} = \sigma_3 \tilde{a}_{\sigma_3},
\end{equation}
where $\sigma_3=\pm 1$ is the eigenvalue. The polarization vector is simply
$\mathbf{n}=\sigma_3 \mathbf{w}$.
In that case, the Stokes parameters in any other MAS $\mathbf{I}'$ are constant; the polarization vector assumes the same form,
$\mathbf{n}'=\sigma_3 \mathbf{w}$.
Here we see again the unique role of the longitudinal Stokes parameter in the polarization.
As a matter of fact, a recent quantum-mechanical analysis \cite{Li16} showed that it is exactly the helicity, the magnitude of the spin.

In the language of quantum mechanics, Eq. (\ref{PV-a}) says that only in one particular MAS can the polarization be represented by the Pauli matrices (\ref{PM}) in the corresponding two-component representation (\ref{QUT2a}) and thus appear as the true intrinsic degree of freedom of the photon. Since the Pauli matrices satisfy the canonical commutation relation,
$[\hat{\sigma}_i, \hat{\sigma}_j]=2i \epsilon_{ijk} \hat{\sigma}_k$,
except for the factor two, it follows that only in one particular MAS can the polarization be canonically quantized.
This fact in turn makes the SV physically observable.

\section{Physical significance of the SV}\label{effect}

\subsection{Observable effects of the SV}

We have seen that the polarization of light beams is their property in the MAS. We have also seen that only in one particular MAS can the polarization be intrinsically independent of the momentum. To explore the physical significance of the SV, it is proper to compare light beams that have the same Stokes parameters in different MAS's.

For clarity, we consider two such beams that are described by the same Jones function (\ref{tildef}) in two different MAS's $\mathbf I$ and $\mathbf{I}'$, respectively.
The first beam that is described in the MAS $\mathbf I$ has vector wavefunction
$\mathbf{f}_\mathbf{I}(\mathbf{k})=\mathbf{a}_\mathbf{I} f(\mathbf{k})$,
where
\begin{equation}\label{vecaI}
    \mathbf{a}_\mathbf{I}=\varpi \tilde{a}
\end{equation}
is its unit vector wavefunction. Correspondingly, the second beam that is described in the MAS $\mathbf{I}'$ has vector wavefunction
$\mathbf{f}_{\mathbf{I}'} (\mathbf{k})=\mathbf{a}_{\mathbf{I}'} f(\mathbf{k})$,
where
\begin{equation}\label{vecaI'}
    \mathbf{a}_{\mathbf{I}'}=\varpi' \tilde{a}.
\end{equation}
Substituting Eq. (\ref{Varpi'}) into Eq. (\ref{vecaI'}) and noticing Eq. (\ref{vecaI}), we have
\begin{equation}\label{a'-a}
    \mathbf{a}_{\mathbf{I}'}
   =\exp [-i(\hat{\mathbf \Sigma} \cdot \mathbf{w}) \Phi] \mathbf{a}_\mathbf{I}.
\end{equation}
This shows that the second beam is related to the first beam by a local rotation about the momentum and is therefore different from the first beam. Specifically, it is different from the first beam only in the vectorial feature.
Indeed, upon using Eqs. (\ref{vecaI}) and (\ref{vecaI'}) and taking Eqs. (\ref{varpi'}) and (\ref{unitarity1}) into account, we find
$\mathbf{a}^\dag_\mathbf{I} \mathbf{a}_{\mathbf{I}'}=\cos \Phi -i \varsigma_3 \sin \Phi$,
which is not equal to unity.
The SV thus has a physical effect in the sense that it acts as the degree of freedom to distinguish the vectorial feature of light beams that have the same Stokes parameters or the same ``polarization''.
This effect can be experimentally observed by detecting the electric vector in accordance with Eq. (\ref{FT}).

\subsection{Application to eigen states of the helicity}

As an important application, we assume that the two beams are both the eigen states of the helicity $\hat{\sigma}_3$ with the same eigenvalue $\sigma_3$ so that their Jones function is given by
$\tilde{f}_{\sigma_3} (\mathbf{k})=\tilde{a}_{\sigma_3} f(\mathbf{k})$.
In this case, the vector wavefunctions of the two beams become
$\mathbf{f}_{\mathbf{I}; \sigma_3} =\mathbf{a}_{\mathbf{I}; \sigma_3} f$
and
$\mathbf{f}_{\mathbf{I}';\sigma_3} =\mathbf{a}_{\mathbf{I}';\sigma_3} f$,
respectively, where
\begin{subequations}
\begin{align}
  \mathbf{a}_{\mathbf{I}; \sigma_3} & =\varpi  \tilde{a}_{\sigma_3}
 =\frac{1}{\sqrt 2}(\mathbf{u} +i \sigma_3 \mathbf{v}),           \label{as3}\\
  \mathbf{a}_{\mathbf{I}';\sigma_3} & =\varpi' \tilde{a}_{\sigma_3}
 =\frac{1}{\sqrt 2}(\mathbf{u}'+i \sigma_3 \mathbf{v}').          \label{as3'}
\end{align}
\end{subequations}
Substituting Eq. (\ref{varpi'}) into Eq. (\ref{as3'}) and using Eqs. (\ref{EE}) and (\ref{as3}), we get
\begin{equation}\label{BP}
    \mathbf{a}_{\mathbf{I}';\sigma_3}
   =\exp (-i \sigma_3 \Phi) \mathbf{a}_{\mathbf{I}; \sigma_3}.
\end{equation}
A comparison of Eq. (\ref{BP}) with Eq. (\ref{a'-a}) shows that the rotation operator on the unit vector wavefunction in this special case reduces to a phase factor that depends on the eigenvalue of the helicity. This means that the vector wavefunctions of the two beams differ only by a helicity-dependent phase.
But as just mentioned, they are not the same at all.

Furthermore, if the two beams are plane waves with the same momentum $\mathbf{k}_0$, that is, if their Jones function is
$\tilde{f}_{\sigma_3,\mathbf{k}_0}=\tilde{a}_{\sigma_3}\delta^3(\mathbf{k}-\mathbf{k}_0)$,
their vector wavefunctions are given by
\begin{subequations}\label{EF-VR}
\begin{align}
  \mathbf{f}_{\mathbf{I};  \sigma_3, \mathbf{k}_0} &
 =\mathbf{a}_{\mathbf{I};  \sigma_3} \delta^3 (\mathbf{k}-\mathbf{k}_0), \label{VWF}  \\
  \mathbf{f}_{\mathbf{I}'; \sigma_3, \mathbf{k}_0} &
 =\mathbf{a}_{\mathbf{I}'; \sigma_3} \delta^3 (\mathbf{k}-\mathbf{k}_0), \label{VWF'}
\end{align}
\end{subequations}
respectively.
These are two plane waves that have the same helicity and the same momentum. From Eqs. (\ref{BP}) and (\ref{FT}) it follows that their electric vectors in position space are different from each other only by a phase. They are therefore hardly distinguishable in classical regime. Unfortunately, they are really treated as the same state in traditional quantum optics \cite{CDG, MW}.
However, it was recently shown in quantum-mechanical regime \cite{Li16} that the helicity-dependent phase difference between the vector wavefunctions (\ref{VWF}) and (\ref{VWF'}) is nothing but a Berry phase \cite{Berry84}. It reflects such a fact that the SV is the degree of freedom to determine the barycenter of the photon of definite helicity and definite momentum. This shows \cite{Li09-2} why the so-called spin Hall effect of light \cite{HK} can be explained \cite{OMN} in terms of the Berry phase.

\section{Conclusions}\label{remark}

In conclusion, we summarize the distinction between the Poincar\'{e} vector (\ref{PV-a}) and the unit vector wavefunction (\ref{QUT1-a}) it characterizes.
Firstly, the former is a real-valued vector. It describes the polarization of light beams in the MAS that is fixed by the SV. The latter is a complex-valued vector. It describes the vectorial feature of light beams in the laboratory reference system.
Secondly, the Cartesian components of the former along the axes of the MAS, the Stokes parameters, can be independent of the momentum. But due to the transversality condition (\ref{TC-a}), the Cartesian components of the latter along the axes of the laboratory reference system cannot be independent of the momentum.
From the standpoint of quantum mechanics, it is the polarization (\ref{PV-a}) in the MAS that is the intrinsic degree of freedom of the photon. This explains not only why the polarization of light beams in any particular MAS is able to characterize their vectorial feature in the laboratory reference system but also why the two plane waves (\ref{EF-VR}) having the same helicity and the same momentum are different from each other.

%\section*{Acknowledgments}

\end{document}